\documentclass[11pt]{article}
\usepackage{graphics,epsfig}
\usepackage[latin1]{inputenc}
\usepackage[english,activeacute]{babel}
\usepackage[]{graphicx}
\usepackage{latexsym}
\usepackage{amsmath, amsthm, amsfonts, amssymb}
\usepackage{verbatim}

\begin{document}

\newtheorem{theo}{Theorem}[section]
\newtheorem{definition}[theo]{Definition}
\newtheorem{lem}[theo]{Lemma}
\newtheorem{prop}[theo]{Proposition}
\newtheorem{coro}[theo]{Corollary}
\newtheorem{exam}[theo]{Example}
\newtheorem{rema}[theo]{Remark}
\newtheorem{example}[theo]{Example}
\newtheorem{principle}[theo]{Principle}
\newcommand{\ninv}{\mathord{\sim}}
\newtheorem{axiom}[theo]{Axiom}

\setcounter{page}{1} 
\setcounter{footnote}{0} 
\renewcommand{\thefootnote}{\fnsymbol{footnote}}

\title{\MakeUppercase{Entities, Identity and the Formal Structure of Quantum Mechanics}}
\author{Christian de Ronde$^{1}$, Graciela Domenech$^{2}$, Federico Holik$^{2}$ and Hector Freytes$^{3,\ 4}$}
\date{}
\renewcommand{\date}{\vspace{-5mm}}
\maketitle \vspace*{-3mm}\relax
\renewcommand{\thefootnote}{\arabic{footnote}}

\noindent \textit{\small $^1$ Center Leo Apostel (CLEA)\\Foundations
of the Exact Sciences (FUND), Brussels Free University,
Krijgskundestraat 33, 1160 Brussels, Belgium}\\
\noindent \textit{\small $^2$ Instituto de Astronom\'{\i}a y
F\'{\i}sica del Espacio (IAFE), Casilla de Correo 67, Sucursal 28,
1428 Buenos Aires, Argentina} \\
\noindent \textit{\small $^3$ Instituto Argentino de Matem\'{a}tica
(IAM), Saavedra 15 - 3er Piso - 1083 Buenos Aires, Argentina}\\
\noindent \textit{\small $^4$ Universit\`{a} degli Studi di
Cagliari, Dipartimento di Scienze Pedagogiche e Filosofiche, Via Is
Mirrionis 1,09123 Cagliari, Italia}

\begin{abstract}
\noindent The concept of individuality in quantum mechanics shows
radical differences from the one used in classical physics. In
particular, it is not possible to consider the fundamental particles
described by quantum theory as individual distinguishable objects.
In this paper we present arguments in favor of quantum
non-individuality, which ---in addition to those based on quantum
statistics--- relate to the Kochen-Specker theorem and the principle
of superposition. Then, we analyze the possibility of referring to
`possible individuals' instead of `actual individuals', and show
that the Modal-Kochen-Specker theorem precludes this
interpretational move.
\end{abstract}

\section{Classical physics as a theory of entities}\label{s:entity}

Using the opportunity of this interdisciplinary conference on the
subject of {\it Structure and Identity} we would like to present a
paper which discusses the physical and philosophical problems
engaged with the interpretation of the formal structure of quantum
mechanics in relation to the notions of entity and identity.

Aristotle relates `identity' to the concept of `entity' via his
formulation of logic. Plato had designed this concept in order to
escape the problem of movement in which Pre-Socratic philosophy had
based itself. The principles of Aristotelian logic, namely, {\it the
existence of objects of knowledge}, {\it the principle of
contradiction} and {\it the principle of identity}, appear thus, as
the conditions of possibility to refer to an entity.\footnote{See
for discussion \cite{VerelstCoecke}.} But as stressed repeatedly by
Martin Heidegger, since Plato,\footnote{We could add Aristotle as
that whom structures the notion of entity beyond philosophy and
constructs the logical structure of it.} philosophy has thought of
Being ---the most radical question regarding philosophy--- in the
very specific terms of an entity, forgetting the question which
guided philosophy in the first place. Alfred North Whitehead also
referred to this aspect of occidental philosophy mentioning with an
ironic glance that: ``The safest general characterization of the
European philosophical tradition is that it consists of a series of
footnotes to Plato'' \cite{Whitehead}. In relation to physics, the
idea of entity played the most radical role in understanding
phenomena, so in this very same sense, as discussed in
\cite{deRondeUQMCD, deRondeOP}, we might say that the history of
classical physics has been confined ---since Plato, but more
specifically in relation to physics, since Aristotle--- to the
history of `physical entities', i.e. particles, waves, fields, etc.

The foundational revolution in mathematics and physics, which took
place in between the end of the 19th century and the beginning of
the 20th, shacked the very cornerstones of classical thought.
Quantum mechanics can be regarded as one of the most definite blows
received by the classical world-view. The impossibility to refer to
an object, an individual, reflected the limitations of the physics
which had been thought until then, a physics which had been founded
on the concept of `entity' and the logical structure of Aristotle.
Expressing the idea that one needs to apply the principle of
identity to every entity, Heidegger states in {\it Der Staz der
Identit\"at} that the principle is a law of Being which states that
identity ---the unit with itself--- pertains to each entity as such.
This principle ---in the form `$a=a$'--- belongs to the axioms of
set theory, which is in turn in the basis of the mathematics in
which physical theories are axiomatized.

We will analyze the compatibility between the notions of entity and
identity and the formal structure of quantum mechanics. To do so, we
assume  that a certain description is determined by a set of
concepts which interrelate and that a concept is defined  by its
inter-relations to other concepts. Identity and individuality are
two main constituters of the notion of entity. In this respect, many
discussions tend to `dissect' the concept of entity, as if, when
tearing apart the concept from its constituents they would be still
allowed to talk about the same concept.\footnote{This is discussed
more deeply in \cite{deRondeNENI}.} But the notion of individual and
that of properties are {\it interdefined} notions. The individual
entity is constituted by the notions of existence, identity and
contradiction (of properties obeying classical logic), these are
notions which interrelate and determine the concept of entity. The
fall of one of them causes the whole architectonic to loose its
foundation and crumble down into pieces.

We shall firstly review the standard analysis of quantum
non-individuality of indistinguishable particles via the quantum
statistics and comment an alternative approach to deal with them
from quasiset theory \cite{dchgk, krause92}, a landscape where the
logical principle of identity has restricted applicability. Then we
present other challenges to individuality, namely those posed by the
Kochen-Specker theorem and the superposition principle. We also link
the discussion on quantum non-individuality to the interpretation of
the Fock-space formalism, and relate it to the philosophical problem
of the one and the many. Finally we discuss, within the modal scheme
if it is achievable to retain, instead of the notion of `actual
individual', at least the notion of `possible individual'.

\section{Individuality in quantum mechanics}

Since the origins of quantum mechanics, problems appeared when
attempts were made to interpret its formalism and experiments. In
particular, a major problem was found in relation to the
individuality of quantum systems. An expression of this problem can
be found in \cite{Sch98a}, where Erwin Schr\"{o}dinger states that:
``[...] we have [...] been compelled to dismiss the idea that [...]
a particle is an individual entity which retains its `sameness'
forever. Quite the contrary, we are now obliged to assert that the
ultimate constituents of matter have no `sameness' at all.'' And
continues: ``I beg to emphasize this and I beg you to believe it: It
is not a question of our being able to ascertain the identity in
some instances and not being able to do so in others. It is beyond
doubt that the question of `sameness', of identity, really and truly
has no meaning.'' Schr\"{o}dinger extended his assertions in
different works, for example:

\begin{quotation}``I mean this: that the elementary particle is not an individual; it
cannot be identified, it lacks `sameness'. The fact is known to
every physicist, but is rarely given any prominence in surveys
readable by nonspecialists. In technical language it is covered by
saying that the particles `obey' a newfangled statistics, either
Einstein-Bose or Fermi-Dirac statistics. [...] The implication, far
from obvious, is that the unsuspected epithet `this' is not quite
properly applicable to, say, an electron, except with caution, in a
restricted sense, and sometimes not at all." E. Schr\"{o}dinger
(\cite{Sch98}, p.197)\end{quotation}

\noindent Schr\"{o}dinger was in both cases specifically referring
to what are called `indistinguishable particles', the subject where
these questions first arose. But it may be claimed that the whole
formal structure of quantum mechanics is in conflict with the notion
of individuals, as we will discuss below.

\subsection{Quantum statistics: Bose-Einstein and Fermi-Dirac}

It is commonly argued that elementary particles (electrons, quarks,
neutrinos, and so on) are nomological objects: their
characteristical properties are fixed by law. Particles of a given
kind ---for example, electrons--- are by definition all exactly
equal to one another. Inside a class, they are indistinguishable.
Thus, following Leibniz, they would be all considered as {\it one
and the same thing}. In such case, one would have to admit that
there is only one electron in the whole world ---something that is
strongly denied by experiment \cite{toraldo di franchia}.

Quantum theory prescribes a new way of counting electrons, photons
and the like. Quanta obey Bose-Einstein and Fermi-Dirac statistics,
in opposition to classical systems, which obey Maxwell-Boltzmann
statistics.\footnote{See \cite{fyk} for a very complete review.}
Consider two indistinguishable particles, for example electrons. If
particles were assumed to possess individuality, they could be
labeled as `electron $1$' and `electron $2$'. Suppose they have to
be arranged in two boxes. If they were objects, then they could
actually be labeled because they may be at least distinguished by
their positions
---such as `being in the left box' or `being in the right box'---
there are four possibilities: both electrons in the left box, both
in the right box, electron $1$ in the left box and electron $2$ in
the right box and viceversa. These `labeled' electrons would obey
Maxwell-Boltzmann statistics and all four possibilities should be
considered. Instead, quantum statistics impose that the two last
options, because there is no way of distinguishing which electron is
$1$ and which one is $2$, have to be considered as one and the same.
If the particles were bosons, there would only be three
possibilities left, while if they were fermions, there would be only
one ---because, in addition, the two first arrangements are
prohibited for fermions by the Pauli exclusion principle. It is
difficult to justify this way of counting the possible arrangements
if particles are assumed to posses individuality and thus, the
possibility of being labeled. This tension between individuality and
quantum statistics is usually given as a justification for the fact
that indistinguishable particles cannot be considered as
individuals. However, the standard quantum formalism is constituted
in the mathematics based on set theory. But how could quantum
non-individuals be collected  in sets when Cantor stated that sets
may be regarded as ``...collections of definite and separate objects
of our intuition or our thought''? In order to face this serious
inconsistency, there have been several proposed approaches to
develop quantum set theories\footnote{See for example \cite{dchg},
\cite{krause92}, \cite{krause96}. There are also other quantum set
theories \cite{DHFoundations}, \cite{dunn}, \cite{fink},
\cite{nishimura}, \cite{schlesinger}, \cite{takeuti} that follow the
suggestion made by von Neumann in relation to quantum logic. } that
acknowledge indistinguishability ``right from the start''. In these
frames, the principle of identity has restricted applicability.

There is, nevertheless, a standard way to deal with the subject of
quantum statistics retaining particle indexation. This is done by
restricting the states available for the particles by imposing that,
if particles are indistinguishable, then they can only access
symmetrized (with respect to particle interchange) states if they
are bosons, or antisymmetrized states if they are fermions. All
other states are prohibited or it is supposed that they are never
realized in nature. This trick of first labeling and then
restricting the available states allows to reproduce quantum
statistics satisfactorily, without dropping particle indexation.
But, as stated in \cite{ReadheadyTeller92}, this procedure can be
criticized, for one should still acknowledge why there appear in the
formalism `prohibited states' which may be thus considered as
``surplus structure''. In fact, there has been considerable work
around the existence of particles (`paraparticles') obeying other
types of statistics (`parastatistics'). In \cite{ReadheadyTeller92}
it is argued that, as long as parastatistics are never observed in
nature, then the usual procedure to construct the set of possible
states ---technically, the labeled tensor product Hilbert space
formalism--- should be replaced by another one that does not index
particles, for example the Fock-space formalism. Though these
formulations are not always equivalent,\footnote{See, for example,
\cite{Schmontology2002}.} Fock-space formalism covers all
experimental facts described by the standard formulation and
satisfies the requirement that particles are not labeled ---so that
no surplus structure appears within it. It is important to remark
however, that ---as Steven French and D\'{e}cio Krause argue in
\cite{fyk}--- the formal construction of the Fock-space does use the
standard set theoretical framework, which presupposes classical
individuality on its foundations. This seems to be, thus, not a
genuine solution.

The above outlined arguments show a deep blundering in the
understanding of the meaning of the quantum formalism. On the one
hand, experiments seem to rule out individuals but, on the other
hand, the mathematics (set theory) used to formulate quantum
mechanics presupposes the notion of individuality. Furthermore, we
agree with Michael Readhead and Paul Teller when they say that:

\begin{quotation}``Interpreters of quantum mechanics largely agree that
classical concepts do not apply without alteration or restriction to
quantum objects. In Bohr's formulation this means that one cannot
simultaneously apply complementary concepts, such as position and
momentum, without restriction. In particular, this means that one
cannot attribute classical, well defined trajectories to quantum
systems. But in a more fundamental respect it would seem that
physicists, including Bohr, continue to think of quantum objects
classically as individual things, capable, at least conceptually, of
bearing labels. It is this presumption and its implications which we
need to understand and critically examine.'' M. Readhead and P.
Teller (\cite{ReadheadyTeller92}, p.202)\end{quotation}

To deal with the indistinguishability of quanta we believe, along
with Krause \cite{Krause2005}, that quantum set theories ---which
incorporate quantum non individuality without using particle
labeling--- deserve further investigation. We also believe that it
is an interesting task to construct a Fock-space formalism using
quasiset theory, thus avoiding intermediate indexations and
incorporating quantum indistinguishability ``right from the start''
\cite{DHK}.

\subsection{Quantum individuals: Kochen-Specker theorem and superpositions}

Up to now, we have discussed individuality and identity in the case
of the statistical properties of indistinguishable quanta. We go now
a step further and claim that the failure of the applicability of
the notion of individuality occurs in a more general frame. Indeed,
it occurs within the whole structure of quantum mechanics.

Let us consider the set $\mathcal{L}$  of physical properties of a
quantum system. The formalism of the theory associates to each
physical magnitude a mathematical object ---an operator, called
``observable'', over the Hilbert space of states of the system---
and the Heisenberg principle states that not all magnitudes may
posses (accurate) values at the same time. This must not be
interpreted as a consequence of our ignorance or of our inexact
procedures to determine them. Only subsets of `compatible'
magnitudes may simultaneously posses values. The indetermination of
the values of incompatible pairs is a matter of principle. In fact,
it is one of the fundamental physical principles from which the
formal structure of the theory may be derived. In mathematical
terms, observables linked by the Heisenberg principle do not commute
and thus, physical magnitudes obey a non-commutative algebra
---technically, the projectors in which they decompose are structured in a modular lattice in
the finite case. This is strongly different from the classical realm
---where they are structured in a Boolean lattice--- and thus there
exist (Boolean) valuations of all propositions about physical
magnitudes. The different algebraic structure of the quantum
properties has as its counterpart the different meaning of the
logical connectives among propositions regarding properties. Thus,
if we naively try to interpret them as classical properties, as
properties `possessed by the system', we are faced to all kind of
no-go theorems that preclude this possibility. Most remarkably is
the Kochen-Specker (KS) theorem \cite{ks} which explicitly shows the
fact that within the formal structure of quantum mechanics, it is
not possible to jointly assign truth values to different
not-disjoint subsets of mutually compatible properties. This is a
very strong impediment  to get an image of quantum systems in some
sense close to classical objects. Continuing with the investigation
in \cite{deRondeOP, deRondeNENI}, we claim that the conclusion which
must be driven from the KS theorem is that the quantum wave function
cannot be conceived in terms of the state of {\it an individual
which possesses properties}. The possible mathematical
representations which expose the quantum wave function from
different basis cannot be interpreted as related to properties which
preexist simultaneously. Thus, the KS theorem shows the
impossibility to unify the different representations in a unique and
singular `whole', in something which can be considered as {\it an
individual}.

However, not even when a single basis ---a particular mathematical
representation to express the state of the system--- is taken into
account we can return so easily to the notion of individual without
contradicting the formalism of the theory. In general,  a quantum
state (expanded in a chosen basis) is a linear combination of the
elements of the basis. This is what is called a {\it superposition}.
Paul Dirac called special attention with respect to its importance:

\begin{quotation}
``The nature of the relationships which the superposition principle
requires to exist between the states of any system is of a kind that
cannot be explained in terms of familiar physical concepts. One
cannot in the classical sense picture a system being partly in each
of two states and see the equivalence of this to the system being
completely in some other state. There is an entirely new idea
involved, to which one must get accustomed and in terms of which one
must proceed to build up an exact mathematical theory, without
having any detailed classical picture." P. Dirac (\cite{Dirac47},
p.12)\end{quotation}

\noindent The idea of regarding a superposition as representing the
state of an individual that possess properties does not
work.\footnote{See also \cite{deRondeOP} for discussion of the
principle of superposition in relation to the notion of entity.} A
superposition is a pondered by complex numbers sum of various states
of the system. Let us suppose that one of them is `up' and the other
is `down' (or `dead' and `alive', as is the case in the famous
example of poor Schr\"{o}dinger's  cat \cite{miau}). We immediately
recognize that both states `up' and `down' cannot be the
simultaneous states of the same individual entity. The relationship
between the states in the superposition is what Dirac points out as
remaining far away from our familiar physical concepts.

According to the principle of contradiction ---that which Aristotle,
Leibniz  and Kant considered as the most certain of all
principles--- everything {\it is} or {\it is not} the case. But, if
we think of the components of the superposition as states which
exist simultaneously, we are faced to a {\it contradiction}. To
escape from it, attempts were made to interpret the superposition as
a mathematical object expressing that the system is in one of its
(unknown) state components. In which one of them, would be
`discovered' by measurement. Measurement would then reveal the
`true' state of the system. But we know this idea is simply wrong as
has been shown through various theorems \cite{mitt}. It is not
possible to provide an {\it ignorance interpretation} of the
elements of a superposition. The elements of the superposition do
not exist in the mode of being of actuality, the only mode of being
which we are accustomed to call {\it real}.

All these considerations confront ourselves, as in the case of
indistinguishable quanta, with serious difficulties to interpret the
formalism in terms of individual entities. If one firmly believes
that such a thing as a {\it superposition} refers to something which
has physical existence
---and this is what quantum mechanics tells us if taken seriously--- it
seems we might be in the need of creating a new way of dealing with
these elements of the quantum formalism, a way which should not
depend on the classical idea of individual entity.

\section{Superposition of the one and the many}

In general, when physicists say that quantum systems have an {\it
undefined} `particle number', it may be understood that the number
of particles varies in time because the system is open or because
particles are created or destroyed. Thus, it may be presupposed that
the state of affairs is such that a definite particle number can be
attributed to the system even when one does not know it at every
instant of time. In this frame, the Fock-space formalism is used
because, within it, superpositions of states corresponding to
different particle numbers are allowed. But the question arises of
how many particles are encompassed in such a superposition state.

\subsection{How many?}

Performing a single measurement over a quantum system does not
allow, as we have already discussed, to attribute the result of the
measurement to a property which the system possessed before the
measurement was performed without giving rise to serious problems.
This is also the case when the said property is the `number of
particles' of the system. Suppose that the state is a superposition
of two elementary states representing `two particles' and `five
particles'. The coefficients in the superposition, i.e. the numbers
that ponder the elementary states in the sum, allow to predict the
probability of obtaining two or five particles (no other result is
allowed) when performing a measurement. Now, suppose that five
particles are detected in a single measurement. We still cannot
attribute this finding to the actual state before the process of
measurement because the number of particles was not definite. The
assertion that this is so because the number of particles may be
varying in time for particles are being constantly created and
destroyed, assumes that at each instant of time the number of
particles is a determined {\it definite value}, a statement
inconsistent with the formalism.

The conclusion that we are forced to derive from the formalism is
that a system in a state that is a superposition of several defined
particle number states has no cardinal. In other words, we claim
that the particle number is {\it indefinite} in the same sense as
any other property when we deal with a superposition. There are only
few particular cases in which the cardinal can be considered to be
definite valued ---when the state is an eigenstate of the particle
number operator--- or in which we can attribute to ignorance the
indetermination in the cardinal
---when the state is a statistical mixture of defined number states.
But, in the general case, the incapability of knowing the particle
number does not come from our ignorance about the system but from
the fact that the cardinal does not exist. It is important to point
out one more time that the so called `particle number' only appears,
in general, after the measurement process is performed. And, as  the
result of a measurement cannot be attributed in general to a
property pertaining to a system, `counting' in quantum mechanics is
qualitatively different from counting the quantity of elements of
classical systems.

In which sense do we talk about quantum systems composed for example
by a single photon? What do we mean when we use the words `single
photon'? Experiments on quantum systems sometimes show corpuscular
features, and this suggests an idea of individuality. This idea is a
base for developing the concept of `particle' and subsequently, the
notion of `particles aggregate'. But these are definite experimental
arrangements which force the appearance of particle characteristics
as a final result of a single process. Experiments are designed to
find out which {\it is} the particle number, but this does not mean
that the resulting number pertains to the system under study. On the
contrary, it refers to the definite {\it process} which takes place
in each measurement. We are not allowed to consider the system as an
aggregate of individuals as if they were simple objects.

\subsection{What would an adequate formalism be like?}

How would the characteristics be of an ontology which is not founded
on entities? It is hard to imagine it, but we can suspect that `to
have a definite number' would not necessarily need to be a principal
characteristic of it. To have a definite number, is something to
which entities are always tied. Thus, it would not be surprising
that a Fock-space formulation of quantum mechanics based on
quasi-set theory would be more adequate than the wave mechanics
formulation. If we accept that physics refers in some sense to
Being, this could be considered as an interesting example about how
we could speak about Being without appealing to entities. Something
which is not an individual entity, needs not to be {\it a one}, nor
{\it a many}. The (historically constructed) notion of `number'
needs not to be applied to it. Therefore, that which is expressed
through a superposition is not a one, nor a many, but
notwithstanding, \emph{it is}. In order to provide a formalism which
comprises the mentioned features we have proposed an alternative
procedure (\cite{DHK}, \cite{dhkk}) that resembles that of the
Fock-space formalism but based on quasiset theory $\mathcal{Q}$
which genuinely avoids artificial labeling.

We briefly review first the main ideas of quasi-set theory
$\mathcal{Q}$ following mainly \cite{Unestudio}. Intuitively
speaking, $\mathcal{Q}$ is obtained by applying ZFU-like
(Zermelo-Fraenkel plus \textit{Urelemente}) axioms to a basic domain
composed of $m$-atoms (the new ingredients that stand for
indistinguishable quanta, and to which the usual concept of identity
does not apply), $M$-atoms and aggregates of them. The theory still
admits a primitive concept of quasi-cardinal, which intuitively
stands for the `quantity' of objects in a collection. This is made
so that certain quasi-sets $x$ (in particular, those whose elements
are q-objects) may have a quasi-cardinal, written $qc(x)$, but not
an associated ordinal.  It is also possible to define a translation
from the language of ZFU into the language of $\mathcal{Q}$ in such
a way so that there is a `copy' of ZFU in $\mathcal{Q}$ (the
`classical' part of $\mathcal{Q}$). In this copy, all the usual
mathematical concepts can be defined (inclusive the concept of
ordinal for the $\mathcal{Q}$-sets, the copy of standard sets in
$\mathcal{Q}$), and the $\mathcal{Q}$-sets turn out to be those
quasi-sets whose transitive closure (this concept is like the usual
one) does not contain $m$-atoms.

In $\mathcal{Q}$, `pure' quasi-sets have only $m$-atoms as elements
(although these elements may be not always indistinguishable from
one another), and to them it is assumed that the usual notion of
identity cannot be applied (the expressions $x=y$ and its negation,
$x \not= y$, are not well formed formulas if either $x$ or $y$ stand
for $m$-atoms). Notwithstanding, there is a primitive relation
$\equiv$ of indistinguishability having the properties of an
equivalence relation, and a defined concept of \textit{extensional
identity}, not holding among $m$-atoms, which has the properties of
standard identity of classical set theories. More precisely, we
write $x =_E y$ ($x$ and $y$ are extensionally identical) iff they
are both qsets having the same elements (that is, $\forall z (z \in
x \longleftrightarrow z \in y)$) or they are both $M$-atoms and
belong to the same qsets (that is, $\forall z (x \in z
\longleftrightarrow y \in z)$). From now on, we shall use the symbol
``='' for the extensional equality, except when explicitly
mentioned.

Since the elements of a quasi-set may have properties (and satisfy
certain formulas), they can be regarded as
\textit{indistinguishable} without turning to be \textit{identical}
(that is, being \textit{the same} object), that is, $x \equiv y$
does not entail $x=y$. Since the relation of equality (and the
concept of identity) does not apply to $m$-atoms, they can also be
thought of as entities devoid of individuality. For details about
$\mathcal{Q}$ and about its historical motivations, see \cite[Chap.\
7]{fyk}.

One of the main features of $\mathcal{Q}$ is its ability to take
into account in `set-theoretical terms' the non observability of
permutations in quantum physics, which is one of the most basic
facts regarding indistinguishable quanta. In standard set theories,
if $w \in x$, then of course $(x - \{w \}) \cup \{z\} = x$ iff $z =
w$. That is, we can `exchange' (without modifying the original
arrangement) two elements iff they are \textit{the same} element, by
force of the axiom of extensionality. But in $\mathcal{Q}$ there is
a theorem guarantying the unobservability of permutations; in other
words,

\begin{theo}
Let $x$ be a finite quasi-set such that $x$ does not contain all
indistinguishable from $z$, where $z$ is an $m$-atom such that $z
\in x$. If $w \equiv z$ and $w \notin x$, then there exists $w'$
such that $(x - z') \cup w' \equiv x$
\end{theo}
Here $z'$ and $w'$ stand for a quasi-set with quasi-cardinal 1 whose
only element is indistinguishable (but not identical) from $z$ and
$w$ respectively.

We outline now the construction of the state space $V_{Q}$ that
respects indistinguishability in all steps working within
$\mathcal{Q}$, mainly following \cite{DHK} and \cite{dhkk}. Let us
consider a quasi-set $\epsilon= \{\epsilon_{i}\}_{i \in I }$, where
$I$ is an arbitrary collection of indexes (this makes sense in the
`classical part' of $\mathcal{Q}$). We take the elements
$\epsilon_{i}$ to represent the eigenvalues of a physical magnitude
of interest. Consider then the quasi-functions $f$ (this concept
generalizes that of function), $f:\epsilon \longrightarrow
\mathcal{F}_{p}$, where $\mathcal{F}_{p}$ is the quasi-set formed of
finite and pure quasi-sets. $f$ is the quasi-set formed of ordered
pairs $\langle \epsilon_{i};x\rangle$ with $\epsilon_{i}\in\epsilon$
and $x\in\mathcal{F}_{p}$. Let us choice these quasi-functions in
such a way that whenever $\langle \epsilon_{i_{k}};x\rangle$ and
$\langle \epsilon_{i_{k'}};y\rangle$ belong to $f$ and $k\neq  k'$,
then $x\cap y=\emptyset$. Let us further assume  that the sum of the
quasi-cardinals of the quasi-sets which appear in the image of each
of these quasi-functions is finite, and then, $qc(x)=0$ for every
$x$ in the image of $f$, except for a finite number of elements of
$\epsilon$. Let us call $\mathcal{F}$ the quasi-set formed of these
quasi-functions. If $\langle x;\epsilon_{i}\rangle$ is a pair of
$f\in\mathcal{F}$, we will interpret that the energy level
$\epsilon_{i}$ has occupation number $qc(x)$. These quasi-functions
will be represented by symbols such as
$f_{\epsilon_{i_{1}}\epsilon_{i_{2}}\ldots\epsilon_{i_{m}}}$ (or by
the same symbol with permuted indexes). This indicates that the
levels $\epsilon_{i_{1}}\epsilon_{i_{2}}\ldots\epsilon_{i_{m}}$ are
occupied. It will be taken as convention that if the symbol
$\epsilon_{i_{k}}$ appears $j$-times, then the level
$\epsilon_{i_{k}}$ has occupation number $j$. The levels that do not
appear have occupation number zero.

It is important to point out that the order of the indexes in
$f_{\epsilon_{i_{1}}\epsilon_{i_{2}}\ldots\epsilon_{i_{n}}}$ has no
meaning at all because up to now, there is no need to define any
particular order in $\epsilon$, the domain of the quasi-functions of
$\mathcal{F}$. Nevertheless, we may define an order in the following
way. For each quasi-function $f\in\mathcal{F}$, let
$\{\epsilon_{i_{1}}\epsilon_{i_{2}}\ldots\epsilon_{i_{m}}\}$ be the
quasi-set formed by the elements of $\epsilon$ such that
$\langle\epsilon_{i_{k}},X\rangle\in f$ and $qc(X)\neq  0$ ($k=
1\ldots m$). We call $supp(f)$ this quasi-set (the \textit{support}
of $f$). Then consider the pair $\langle o,f\rangle$, where $o$ is a
bijective quasi-function
$o:\{\epsilon_{i_{1}}\epsilon_{i_{2}}\ldots\epsilon_{i_{m}}\}\longrightarrow
\{1,2,\ldots,m\}.$ Each of these quasi-functions $o$ define an order
on $supp(f)$. For each $f\in\mathcal{F}$, if $qc(supp(f))= m$, then,
there are $m!$ orderings. Then, let $\mathcal{O}\mathcal{F}$ be the
quasi-set formed by all the pairs $\langle o,f\rangle$, where
$f\in\mathcal{F}$ and $o$ is a a particular ordering in $supp(f)$.
Thus, $\mathcal{O}\mathcal{F}$ is the quasi-set formed by all the
quasi-functions of $\mathcal{F}$ with ordered support. For this
reason,
$f_{\epsilon_{i_{1}}\epsilon_{i_{2}}\ldots\epsilon_{i_{n}}}\in
\mathcal{O}\mathcal{F}$ refers to a quasifunction $f\in\mathcal{F}$
and a special ordering of
$\{{\epsilon_{i_{1}}\epsilon_{i_{2}}\ldots\epsilon_{i_{n}}}\}$. For
the sake of simplicity, we will use the same notation as before. But
now the order of the indexes {\it is meaningful}. It is also
important to remark, that the order on the indexes must not be
understood as a labeling of particles, for it easy to check, as
above, that the permutation of particles does not give place to a
new element of $\mathcal{O}\mathcal{F}$. This is so because a
permutation of particles operating on a pair $\langle
o,f\rangle\in\mathcal{O}\mathcal{F}$ will not change $f$, and so,
will not alter the ordering. We will use the elements of
$\mathcal{O}\mathcal{F}$ later, when we deal with fermions.

A linear space structure is required to adequately represent quantum
states. To equip $\mathcal{F}$ and $\mathcal{OF}$ with such a
structure, we need to define two operations ``$\star$" and ``$+$", a
product by scalars  and an addition of their elements, respectively.
Call $C$ the collection of quasi-functions which assign to every
$f\in \mathcal{F}$ (or $f\in \mathcal{O}\mathcal{F}$) a complex
number (again, built in the `classical part' of $\mathcal{Q}$). That
is, a quasi-function $c\in C$ is a collection of ordered pairs
$\langle f;\lambda\rangle$, where $f\in \mathcal{F}$ (or $f\in
\mathcal{O}\mathcal{F}$) and $\lambda$ a complex number. Let $C_{0}$
be the subset of $C$ such that, if $c\in C_0$, then $c(f)=0$ for
almost every $f\in \mathcal{O}\mathcal{F}$ (i.e., $c(f)=0$ for every
$f\in \mathcal{O}\mathcal{F}$ except for a finite number of
quasi-functions). We can define in $C_{0}$ a sum and a product by
scalars in the same way as it is usually done with functions as
follows:
\begin{definition}
Let $\alpha$, $\beta$ and $\gamma$ $\in \mathcal{C}$, and $c$,
$c_{1}$ and $c_{2}$ be  quasi-functions of $C_{0}$, then
$$(\gamma\ast c)(f) := \gamma(c(f))\ \ and \ \
(c_{1}+c_{2})(f) :=  c_{1}(f) + c_{2}(f)$$
\end{definition}
\noindent The quasi-function $c_{0}\in C_{0}$ such that $c_{0}(f)=
0$, for any $f\in F$, acts as the null element of the sum, for $
(c_{0}+c)(f)= c_{0}(f)+c(f)= 0+c(f)= c(f), \forall f.$ With the sum
and the multiplication by scalars defined above we have that
$(C_{0},+,\ast)$ is a complex vector space. Each one of the
quasi-functions of $C_{0}$ should be interpreted in the following
way: if $c\in C_{0}$ (and $c\neq c_{0}$), let $f_{1}$, $f_{2}$,
$f_{3}$,$\ldots$, $f_{n}$ be the only functions of $C_{0}$ which
satisfy $c(f_{i})\neq  0$ ($i= 1,\ldots,n$). These quasi-functions
exist because, as we have said above, the quasi-functions of $C_{0}$
are zero except for a finite number of quasi-functions of
$\mathcal{F}$. If $\lambda_{i}$ are complex numbers which satisfy
that $c(f_{i})= \lambda_{i}$ ($i= 1,\ldots,n$), we will make the
association
$c\approx(\lambda_{1}f_{1}+\lambda_{2}f_{2}+\cdots+\lambda_{n}f_{n})$.
The symbol $\approx$ must be understood in the sense that we use
this notation to represent the quasi-function $c$. The idea is that
the quasi-function $c$ represents the pure state which is a linear
combination of the states represented by the quasi-functions $f_{i}$
according to the interpretation given above.

In order to calculate probabilities and mean values, we have to
introduce a scalar product, in fact two of them:  $\circ$ for bosons
and $\bullet$ for fermions, thus obtaining two (normed) vector
spaces $(V_{Q},\ \circ )$ and  $(V_{Q},\ \bullet )$ :
\begin{definition}
Let $\delta_{ij}$ be the Kronecker symbol and
$f_{\epsilon_{i_{1}}\epsilon_{i_{2}}\ldots\epsilon_{i_{n}}}$ and
$f_{\epsilon_{i'_{1}}\epsilon_{i'_{2}}\ldots\epsilon_{i'_{m}}}$ two
basis vectors, then
$$
f_{\epsilon_{i_{1}}\epsilon_{i_{2}}\ldots\epsilon_{i_{n}}}\circ
f_{\epsilon_{i'_{1}}\epsilon_{i'_{2}}\ldots\epsilon_{i'_{m}}} :=
\delta_{nm}\sum_{p}\delta_{i_{1}pi'_{1}}\delta_{i_{2}pi'_{2}}\ldots\delta_{i_{n}pi'_{n}}
$$
The sum is extended over all the permutations of the set
$i'=(i'_{1},i'_{2},\ldots,i'_{n})$ and for each permutation $p$,
$pi'=(pi'_{1},pi'_{2},\ldots,pi'_{n})$.
\end{definition}
This product can be easily extended over linear combinations.
\begin{definition}
Let $\delta_{ij}$ be the Kronecker symbol,
$f_{\epsilon_{i_{1}}\epsilon_{i_{2}}\ldots\epsilon_{i_{n}}}$ and
$f_{\epsilon_{i'_{1}}\epsilon_{i'_{2}}\ldots\epsilon_{i'_{m}}}$ two
basis vectors, then
$$f_{\epsilon_{i_{1}}\epsilon_{i_{2}}\ldots\epsilon_{i_{n}}}\bullet
f_{\epsilon_{i'_{1}}\epsilon_{i'_{2}}\ldots\epsilon_{i'_{m}}} :=
\delta_{nm}\sum_{p}\sigma_{p}\delta_{i_{1}pi'_{1}}\delta_{i_{2}pi'_{2}}\ldots\delta_{i_{n}pi'_{n}}$$
where: $s^{p}=+1$ if $p$ is even and $s^{p}= -1$ if $p$ is odd.
\end{definition}
The result of this second product $\bullet$ is an antisymmetric sum
of the indexes which appear in the quasi-functions. In order that
the product is well defined, the quasi-functions must belong to
$\mathcal{O}\mathcal{F}$. Once this product is defined over the
basis functions, we can extend it to linear combinations, in a
similar way as for bosons. If the occupation number of a product is
more or equal than two, then the vector has null norm. As it is a
vector of null norm, the product of this vector with any other
vector of the space would yield zero, and thus the probability of
observing a system in a state like it vanishes. This means that we
can add to any physical state an arbitrary linear combination of
null norm vectors for they do not contribute to the scalar product,
which is the meaningful quantity.

With these tools and using the language of $\mathcal{Q}$, the
formalism of QM may be completely rewritten giving a straightforward
answer to the problem of giving a formulation of QM in which
intrinsical indistinguishability is taken into account from the
beginning, without artificially introducing extra postulates. We
make the following association in order to turn the notation similar
to that of the standard formalism. For each quasi-function
$f_{\epsilon_{i_{1}}\epsilon_{i_{2}}\ldots \epsilon_{i_{n}}}$ of the
quasi-sets $\mathcal{F}$ or $\mathcal{O}\mathcal{F}$ constructed
above, we will write $\alpha
f_{\epsilon_{i_{1}}\epsilon_{i_{2}}\ldots\epsilon_{i_{n}}}:=
\alpha|\epsilon_{i_{1}}\epsilon_{i_{2}}\ldots \epsilon_{i_{n}})$
 with the obvious corresponding generalization for linear
combinations.  Once normalized to unity, the states constructed
using $\mathcal{Q}$, are equivalent to the symmetrized vectors for
bosonic states and we have shown that commutation relations
equivalent to the usual ones hold, thus being both formulations
equivalent for bosons.

For fermions, there are some subtleties involved in the
construction. First of all, let us recall the action of the creation
operator $c^{\dag}_\alpha$: let $\zeta$ represent a collection of
indexes with non null occupation number, then $
C^{\dag}_\alpha|\zeta)=|\alpha\zeta)$. If $\alpha$ was already in
the collection $\zeta$, then $|\alpha\zeta)$ is a vector with null
norm. As said above, to have null norm implies that
$(\psi|\alpha\zeta)=0$ for all $|\psi)$. Moreover, if a linear
combination of null norm vectors were added to the vector
representing the state of a system, this addition would not give
place to observable results because the terms of null norm do not
contribute to the mean values or to the probabilities. In order to
express this situation, we define the following relation:
\begin{definition}\label{e:chirimbolo} Two vectors $|\varphi)$ and $|\psi)$
are similar (and we will write $ |\varphi)\cong|\psi)$)) if the
difference between them is a linear combination of null norm
vectors.
\end{definition}
With all of this, it is straightforward to demonstrate the
equivalence of the anticommutation relations in $V_{Q}$ and in the
standard Fock-space. Thus, we can conclude that both formulations
are equivalent also for fermions. To avoid particle labeling in the
expressions for observables, in Fock-space formalism they are
written in terms of creation and annihilation operators.

\section{Modal interpretations of quantum mechanics}

We have argued against forcing the interpretation of the quantum
formalism in terms of  actual individuals.  In view of the
difficulties posed by the theory to give place to actual entities,
modal interpretations of QM attempt to consider the role of {\it
possible properties} in the orthodox quantum formalism giving place
to a consistent discourse about {\it possible entities}.

Modal interpretations have continued the footprints left by Niels
Bohr, Max Born, Werner Heisenberg and Wolfgang Pauli and continued
the path on the lines drawn by Bas van Fraassen, Simon Kochen,
Dennis Dieks and many others, searching for the different
possibilities of interpreting the formalism of the theory.\footnote{
See for example \cite{Dieks88, Kochen85, VF81, VF91}.} Modal
interpretations may be thought to be a study of the constraints
under which one is able to talk a consistent classical discourse
without contradiction with the quantum formalism. Following the
general outline provided in \cite{DFRStudies} one might state in
general terms that a modal interpretation is best characterized by
the following points:

\begin{enumerate}
\item
One of the most significant features of modal interpretations is
that they {\it stay close to the standard formulation}. Following
van Fraassen's recommendation, one needs to learn from the formal
structure of the theory in order to develop an interpretation. This
is different from many attempts which presuppose an ontology and
then try to fit it into the formalism.

\item
Modal interpretations are {\it non-collapse interpretations}. The
evolution is always given by the Schr\"odinger equation of motion
and the collapse of the wave function is nothing but the path from
the possible to the actual, it is not considered a physical process.

\item
Modal interpretations {\it ascribe possible properties to quantum
systems}. The property ascription depends on the states of the
systems and applies regardless of whether or not measurements are
performed.  There is a distinction between the level of possibility
and that of actuality which are related through an interpretational
rule. (Technically, in addition to actual properties interpreted in
the orthomodular lattice ${\cal L}$ of quantum logic, there is a set
of possible properties $\diamond {\cal L}$ which may be regarded as
constituting the center of the enriched lattice.)

\item
Modality {\it is not interpreted in terms of ignorance}. There is no
ignorance interpretation of the probability distribution assigned to
the physical properties. The state of the system determines all
there is to know. For modal interpretations there is no such thing
as `hidden variables' from which we could get more information. One
can formulate a KS theorem for modalities which expresses the
irreducible contextual character of quantum mechanics even in the
case of enriching its language with a modal operator.

\end{enumerate}

\subsection{`Possible' individuals?}

Modal interpretations intend to discuss about systems with
properties going beyond the instrumentalist positions which only
talk about measurement outcomes. However, it is not obvious which
properties  can be considered as possessing definite values. In
particular there are several no-go theorems which restrict this
possibility \cite{Bacciagaluppi95, Vermaas97}.

Still today there is a tension in modal interpretations which has
not been solved. Although Bacciagaluppi claims that ``despite the
name, the modal interpretation in the version of Vermaas and Dieks
is a theory about actualities'' (\cite{Bacciagaluppi96}, p.74),
Dieks still seems to present a different position:

\begin{quotation} ``[...] according to
modal interpretations the quantum formalism does not tell us what
actually is the case in the physical world, but rather provides us
with a list of possibilities and their probabilities. The modal
viewpoint is therefore that quantum theory is about what may be the
case, in philosophical jargon, quantum theory is about modalities.''
D. Dieks  (\cite{Dieks05})\end{quotation}

\noindent It might be thought that such interpretation of quantum
mechanics in terms of modalities opens the door to a new way to
refer to individuals, not in terms of actuality but rather in terms
of possibilities. Thus, one might speak about `possible individuals'
instead of `actual individuals'. At first sight, one might think it
is still achievable to recover the notion of individuality and
identity in quantum mechanics if one is careful enough to remain
within the realm of possibility. But, once again, the quantum
formalism is ready to stop any move which intends to constrain its
interpretation within the notion of individuality. In fact, a
theorem has been developed which precludes to consistently interpret
the formalism in terms of `possible individuals'.

\subsection{Limits of modality: the MKS theorem}

In order to stay away from inconsistencies when speaking about
properties which pertain to the system, one must acknowledge the
limitations imposed by the KS theorem. To do so, modal
interpretations assign to the system only a set of definite
properties. This is not achievable when talking about properties
which pertain to different contexts (see for discussion
\cite{BV99}).

At first sight it might seem paradoxical that, even though modal
interpretations of quantum mechanics talk about modalities, KS
theorem refers to actual values of physical properties. Elsewhere,
and following the line of thought of quantum logic, we have
investigated the question whether KS theorem has something to say
about possibility and its relation to actuality \cite{DFRAnnalen,
DFRIJTP}. The answer was provided via a characterization of the
relations between actual and possible properties pertaining to
different contexts. By applying algebraic and topological tools we
studied the structure of the orthomodular lattice of actual
propositions enriched with modal propositions. Let us briefly recall
the results.  As usual, given a proposition about the system, it is
possible to define a context from which one can predicate with
certainty about it (and about a set of propositions that are
compatible with it) and predicate probabilities about the other
ones. This is to say that one may predicate truth or falsity of all
possibilities at the same time, i.e. possibilities allow an
interpretation in a classical system of propositions. In order to
describe rigorously the formalism which allows to capture all
propositions in a single structure, let ${\cal L}$ be an
orthomodular lattice. Given $a, b, c$ in $L$, we write: $(a,b,c)D$\
\   iff $(a\lor b)\land c = (a\land c)\lor (b\land c)$;
$(a,b,c)D^{*}$ iff $(a\land b)\lor c = (a\lor c)\land (b\lor c)$ and
$(a,b,c)T$\ \ iff $(a,b,c)D$, (a,b,c)$D^{*}$ hold for all
permutations of $a, b, c$. An element $z \in {\cal L}$ is called
{\it central} iff for all elements $a,b\in L$ we have\ $(a,b,z)T$.
We denote by $Z({\cal L})$ the set of all central elements of ${\cal
L}$ and it is called the {\it center} of ${\cal L}$. $Z({\cal L})$
is a Boolean sublattice of $L$ \cite[Theorem 4.15]{mm}.

Let $P$ be a proposition about a system and consider it as an
element of an orthomodular lattice ${\cal L}$. If we refer with
$\Diamond P$ to the possibility of $P$, then $\Diamond P$ will be a
central element of ${\cal L}$. This interpretation of the
possibility in terms of the Boolean algebra of central elements of
${\cal L}$ reflects the fact that one can simultaneously predicate
about all possibilities. This is so because it is always possible to
establish Boolean homomorphisms of the form $v:Z({\cal L})
\rightarrow {\bf 2}$. Therefore, the key idea is to expand the
orthomodular structure in such a way to include propositions about
possibility. This expansion is performed in the following way: If
$P$ is a proposition about the system and $P$ occurs, then it is
trivially possible that $P$ occurs. This is expressed as $P \leq
\Diamond P$. In fact, to assume an actual property and a complete
set of properties that are compatible with it determines a context
in which the classical discourse holds. Classical consequences that
are compatible with it, for example probability assignments to the
actuality of other propositions, shear the classical frame. These
consequences are the same ones as those which would be obtained by
considering the original actual property as a possible property.
This is interpreted as, if $P$ is a property of the system,
$\Diamond P$ is the smallest central element greater than $P$. With
these tools, we are able to give an extension of the orthomodular
structure by adding a possibility operator that fulfills the
mentioned requirements. More precisely, the extension is a class of
algebras, called Boolean saturated orthomodular lattices, that
admits the orthomodular structure as a reduct and we demonstrate
that they are a variety, i.e., definable by equations.

If ${\cal L}$ is an orthomodular lattice and ${\cal L}^{\Diamond}$ a
Boolean saturated orthomodular one such that ${\cal L}$ can be
embedded in ${\cal L}^{\Diamond}$, we say that ${\cal L}^{\Diamond}$
is a modal extension of ${\cal L}$. Given  ${\cal L}$ and a modal
extension ${\cal L}^{\Diamond}$,  we define the {\it possibility
space} as the subalgebra of ${\cal L}^{\Diamond}$ generated by $
\{\Diamond P : P \in {\cal L} \} $. We denote by $\Diamond {\cal L}$
this space and it may be proved that it is a Boolean subalgebra of
the modal extension. The possibility space represents the modal
content added to the discourse about properties of the system.

Within this frame, the actualization of a possible property acquires
a rigorous meaning. Let ${\cal L}$ be an orthomodular lattice,
$(W_i)_{i \in I}$ the family of Boolean sublattices of ${\cal L}$
and ${\cal L}^\Diamond$ a modal extension of $\cal L$. If $f:
\Diamond {\cal L} \rightarrow {\bf 2}$ is a Boolean homomorphism, an
actualization compatible with  $f$ is a global valuation $(v_i: W_i
\rightarrow {\bf 2})_{i\in I}$ such that $v_i\mid W_i \cap \Diamond
{\cal L} = f\mid W_i \cap \Diamond {\cal L} $ for each $i\in I$.

{\it Compatible actualizations} represent the passage from
possibility to actuality. When taking into account compatible
actualizations from different contexts, the following KS theorem for
modalities can be proved \cite{DFRAnnalen}:

\begin{theo}\label{ksm}
Let $\cal L$ be an orthomodular lattice. Then $\cal L$ admits a
global valuation iff for each possibility space there exists a
Boolean homomorphism  $f: \Diamond {\cal L} \rightarrow {\bf 2}$
that admits  a compatible actualization.\qed
\end{theo}

The modal KS (MKS) theorem shows that no enrichment of the language
about actual properties results in something close to a classical
image. The conclusion which can be derived from the MKS theorem is
that the formalism of quantum mechanics does not only deny the
possibility of talking about an `actual entity', but even the term
`possible entity' remains a meaningless notion within its domain of
discourse.

\section{Departing from Plato's footnotes}

Nietzsche and Heidegger claim that the tradition initiated by
Aristotle and followed by the rest of the subsequent occidental
philosophers of thinking Being in terms of entities needs to be
criticized. From the landscape of quantum mechanics we may argue
that all investigations point in the direction of its
incompatibility with the notion of entity. First, a quantum system
cannot be thought in terms of the unity of its properties because
there always exist incompatible properties and besides, it is
inconsistent to say that properties pertain to it, not even those
measured properties. This is also the fact when the brought up
property is the number of particles in the system. Furthermore, it
is not allowed to consider `possible entities', because they fall
under analogous criticisms to those of actual ones. Last in our
series of examples, indistinguishable quanta show similar lacking of
individuality when being considered in a collection, which cause a
different statistical behavior than that of classical aggregates.
Perhaps, this is the time in physics in which we need to abandon the
Aristotelian tradition of thinking the world in terms of entities,
in order to make a cogent picture of quantum phenomena.

\section*{Acknowledgements}

We wish to thank the organizers of the conference and specially to
Karin Verelst and Wim Christiaens for their invitation. G. Domenech
is fellow of the Consejo Nacional de Investigaciones
Cient\'{\i}ficas y T\'ecnicas (CONICET). This work was partially
supported by the following grants: PICT 04-17687 (ANPCyT), PIP N$^o$
6461/05 (CONICET) and UBACyT N$^o$ X081 and Projects of the Fund for
Scientific Research Flanders G.0362.03 and G.0452.04.

\end{document}